\documentclass[12pt]{article}
\newcommand{\nc}{\newcommand}
\nc{\renc}{\renewcommand}

\usepackage{calc}
\usepackage{ifthen}

\nc{\half}{{\textstyle{1\over2}}}
\nc{\etal}{\mbox{\it et al. }}
\nc{\ie}{{\it i.e.}}
\nc{\eg}{{\it e.g.}}

\renc{\thefootnote}{\arabic{footnote}}
\nc{\capt}[1]{{\bf Figure.} {\small\sl #1}}

\nc{\eqs}[2]{\mbox{Eqs.~(\ref{#1},\,\ref{#2})}}
\nc{\eq}[1]{\mbox{Eq.~(\ref{#1})}}

\nc{\figs}[2]{\mbox{Figs.~(\ref{#1},\,\ref{#2})}}
\nc{\fig}[1]{\mbox{Fig~.(\ref{#1})}}

\nc{\tag}[1]{\label{#1} \marginpar{{\footnotesize #1}}}
\nc{\mtag}[1]{\label{#1} \mbox{\marginpar{{\footnotesize #1}}}}
\renc{\baselinestretch}{1.5}
\newlength{\overeqskip}
\newlength{\undereqskip}
\nc{\be}[1]{\begin{equation} \mbox{$\label{#1}$}}
\nc{\bea}[1]{\begin{eqnarray} \mbox{$\label{#1}$}}
\nc{\Section}[2]{\section{#2}\label{#1}}
\nc{\Bibitem}[1]{\bibitem{#1}}
\nc{\Label}[1]{\label{#1}}

\nc{\eea}{\vspace{\undereqskip}\end{eqnarray}}
\nc{\ee}{\vspace{\undereqskip}\end{equation}}
\nc{\bdm}{\begin{displaymath}}
\nc{\edm}{\end{displaymath}}
\nc{\dpsty}{\displaystyle}
\nc{\bc}{\begin{center}}
\nc{\ec}{\end{center}}
\nc{\ba}{\begin{array}}
\nc{\ea}{\end{array}}
\nc{\bab}{\begin{abstract}}
\nc{\eab}{\end{abstract}}
\nc{\btab}{\begin{tabular}}
\nc{\etab}{\end{tabular}}
\nc{\bit}{\begin{itemize}}
\nc{\eit}{\end{itemize}}
\nc{\ben}{\begin{enumerate}}
\nc{\een}{\end{enumerate}}
\nc{\bfig}{\begin{figure}}
\nc{\efig}{\end{figure}}
\nc{\arreq}{&\!=\!&}
\nc{\arrmi}{&\!-\!&}
\nc{\arrpl}{&\!+\!&}
\nc{\arrap}{&\!\!\!\approx\!\!\!&}
\nc{\non}{\nonumber\\*}
\nc{\align}{\!\!\!\!\!\!\!\!&&}

\def\lsim{\; \raise0.3ex\hbox{$<$\kern-0.75em
      \raise-1.1ex\hbox{$\sim$}}\; }
\def\gsim{\; \raise0.3ex\hbox{$>$\kern-0.75em
      \raise-1.1ex\hbox{$\sim$}}\; }
\nc{\DOT}{\hspace{-0.08in}{\bf .}\hspace{0.1in}}
\nc{\Laada}{\hbox {$\sqcap$ \kern -1em $\sqcup$}}
\nc\loota{{\scriptstyle\sqcap\kern-0.55em\hbox{$\scriptstyle\sqcup$}}}
\nc\Loota{{\sqcap\kern-0.65em\hbox{$\sqcup$}}}
\nc\laada{\Loota}
\nc{\qed}{\hskip 3em \hbox{\BOX} \vskip 2ex}

\nc{\real}{{\rm I \! R}}
\nc{\Z}{{\sf Z \!\!\! Z}}
\nc{\complex}{{\rm C\!\!\! {\sf I}\,\,}}
\def\bigid{\leavevmode\hbox{\small1\kern-3.8pt\normalsize1}}
\def\id{\leavevmode\hbox{\small1\kern-3.3pt\normalsize1}}
\nc{\slask}{\!\!\!/}
\nc{\bis}{{\prime\prime}}
\nc{\pa}{\partial}
\nc{\na}{\nabla}
\nc{\ra}{\rangle}
\nc{\la}{\langle}
\nc{\goto}{\rightarrow}
\nc{\swap}{\leftrightarrow}

\nc{\EE}[1]{ \mbox{$\cdot10^{#1}$} }
\nc{\abs}[1]{\left|#1\right|}
\nc{\at}[2]{\left.#1\right|_{#2}}
\nc{\norm}[1]{\|#1\|}
\nc{\abscut}[2]{\Abs{#1}_{\scriptscriptstyle#2}}
\nc{\vek}[1]{{\rm\bf #1}}
\nc{\integral}[2]{\int\limits_{#1}^{#2}}
\nc{\inv}[1]{\frac{1}{#1}}
\nc{\dd}[2]{{{\partial #1}\over{\partial #2}}}
\nc{\ddd}[2]{{{{\partial}^2 #1}\over{\partial {#2}^2}}}
\nc{\dddd}[3]{{{{\partial}^2 #1}\over
        {\partial #2 \partial #3}}}
\nc{\dder}[2]{{{d #1}\over{d #2}}}
\nc{\ddder}[2]{{{d^2 #1}\over{d {#2}^2}}}
\nc{\dddder}[3]{{d^2 #1}\over
        {d #2 d #3}}
\nc{\dx}[1]{d\,^{#1}x}
\nc{\dy}[1]{d\,^{#1}y}
\nc{\dz}[1]{d\,^{#1}z}
\nc{\dl}[1]{\frac{d\,^{#1}l}{(2\pi)^{#1}}}
\nc{\dk}[1]{\frac{d\,^{#1}k}{(2\pi)^{#1}}}
\nc{\dq}[1]{\frac{d\,^{#1}q}{(2\pi)^{#1}}}

\nc{\cc}{\mbox{$c.c.$ }}
\nc{\hc}{\mbox{$h.c.$ }}
\nc{\cf}{cf.\ }
\nc{\erfc}{{\rm erfc}}
\nc{\Tr}{{\rm Tr\,}}
\nc{\tr}{{\rm tr\,}}
\nc{\pol}{{\rm pol}}
\nc{\sign}{{\rm sign}}
\nc{\bfT}{{\bf T }}

\def\GeV{{\rm\ GeV}}
\def\MeV{{\rm\ MeV}}

\def\TeV{{\rm\ TeV}}

\nc{\cA}{{\cal A}}
\nc{\cB}{{\cal B}}
\nc{\cD}{{\cal D}}
\nc{\cE}{{\cal E}}
\nc{\cG}{{\cal G}}
\nc{\cH}{{\cal H}}
\nc{\cL}{{\cal L}}
\nc{\cO}{{\cal O}}
\nc{\cT}{{\cal T}}
\nc{\cN}{{\cal N}}
\nc{\rvac}[1]{|{\cal O}#1\rangle}
\nc{\lvac}[1]{\langle{\cal O}#1|}
\nc{\rvacb}[1]{|{\cal O}_\beta #1\rangle}
\nc{\lvacb}[1]{\langle{\cal O}_\beta #1 |}
\nc{\bb}{\bar{\beta}}
\nc{\bt}{\tilde{\beta}}
\nc{\ctH}{\tilde{\cal H}}
\nc{\chH}{\hat{\cal H}}
\nc{\1}{\aa}
\nc{\2}{\"{a}}
\nc{\3}{\"{o}}
\nc{\4}{\AA}
\nc{\5}{\"{A}}
\nc{\6}{\"{O}}
\nc{\al}{\alpha}
\nc{\g}{\gamma}
\nc{\Del}{\Delta}
\nc{\e}{\epsilon}
\nc{\eps}{\epsilon}
\nc{\lam}{\lambda}
\nc{\om}{\omega}
\nc{\Om}{\Omega}
\nc{\ve}{\varepsilon}
\nc{\mn}{{\mu\nu}}
\nc{\vp}{\varphi}

\nc{\advp}[3]{{\it  Adv.\ in\ Phys.\ }{{\bf #1} {(#2)} {#3}}}
\nc{\annp}[3]{{\it  Ann.\ Phys.\ (N.Y.)\ }{{\bf #1} {(#2)} {#3}}}
\nc{\apl}[3]{{\it  Appl. Phys. Lett. }{{\bf #1} {(#2)} {#3}}}
\nc{\apj}[3]{{\it  Ap.\ J.\ }{{\bf #1} {(#2)} {#3}}}
\nc{\apjl}[3]{{\it  Ap.\ J.\ Lett.\ }{{\bf #1} {(#2)} {#3}}}
\nc{\app}[3]{{\it Astropart.\ Phys.\ }{{\bf #1} {(#2)} {#3}}}
\nc{\cmp}[3]{{\it  Comm.\ Math.\ Phys.\ }{{ \bf #1} {(#2)} {#3}}}
\nc{\cqg}[3]{{\it  Class.\ Quant.\ Grav.\ }{{\bf #1} {(#2)} {#3}}}
\nc{\epl}[3]{{\it  Europhys.\ Lett.\ }{{\bf #1} {(#2)} {#3}}}
\nc{\ijmp}[3]{{\it Int.\ J.\ Mod.\ Phys.\ }{{\bf #1} {(#2)} {#3}}}
\nc{\ijtp}[3]{{\it Int.\ J.\ Theor.\ Phys.\ }{{\bf #1} {(#2)} {#3}}}
\nc{\jmp}[3]{{\it  J.\ Math.\ Phys.\ }{{ \bf #1} {(#2)} {#3}}}
\nc{\jpa}[3]{{\it  J.\ Phys.\ A\ }{{\bf #1} {(#2)} {#3}}}
\nc{\jpc}[3]{{\it  J.\ Phys.\ C\ }{{\bf #1} {(#2)} {#3}}}
\nc{\jap}[3]{{\it J.\ Appl.\ Phys.\ }{{\bf #1} {(#2)} {#3}}}
\nc{\jpsj}[3]{{\it J.\ Phys.\ Soc.\ Japan\ }{{\bf #1} {(#2)} {#3}}}
\nc{\lmp}[3]{{\it Lett.\ Math.\ Phys.\ }{{\bf #1} {(#2)} {#3}}}
\nc{\mpl}[3]{{\it  Mod.\ Phys.\ Lett.\ }{{\bf #1} {(#2)} {#3}}}
\nc{\ncim}[3]{{\it  Nuov.\ Cim.\ }{{\bf #1} {(#2)} {#3}}}
\nc{\np}[3]{{\it  Nucl.\ Phys.\ }{{\bf #1} {(#2)} {#3}}}
\nc{\npps}[3]{{\it  Nucl.\ Phys.\ Proc.\ Suppl.\ }{{\bf #1} {(#2)} {#3}}}
\nc{\pr}[3]{{\it Phys.\ Rev.\ }{{\bf #1} {(#2)} {#3}}}
\nc{\pra}[3]{{\it  Phys.\ Rev.\ A\ }{{\bf #1} {(#2)} {#3}}}
\nc{\prb}[3]{{\it  Phys.\ Rev.\ B\ }{{{\bf #1} {(#2)} {#3}}}}
\nc{\prc}[3]{{\it  Phys.\ Rev.\ C\ }{{\bf #1} {(#2)} {#3}}}
\nc{\prd}[3]{{\it  Phys.\ Rev.\ D\ }{{\bf #1} {(#2)} {#3}}}
\nc{\prl}[3]{{\it Phys.\ Rev.\ Lett.\ }{{\bf #1} {(#2)} {#3}}}
\nc{\pl}[3]{{\it  Phys.\ Lett.\ }{{\bf #1} {(#2)} {#3}}}
\nc{\prep}[3]{{\it Phys.\ Rep.\ }{{\bf #1} {(#2)} {#3}}}
\nc{\prsl}[3]{{\it Proc.\ R.\ Soc.\ London\ }{{\bf #1} {(#2)} {#3}}}
\nc{\ptp}[3]{{\it  Prog.\ Theor.\ Phys.\ }{{\bf #1} {(#2)} {#3}}}
\nc{\ptps}[3]{{\it  Prog\ Theor.\ Phys.\ suppl.\ }{{\bf #1} {(#2)} {#3}}}
\nc{\physa}[3]{{\it  Physica\ A\ }{{\bf #1} {(#2)} {#3}}}
\nc{\physb}[3]{{\it  Physica\ B\ }{{\bf #1} {(#2)} {#3}}}
\nc{\phys}[3]{{\it Physica\ }{{\bf #1} {(#2)} {#3}}}
\nc{\rmp}[3]{{\it  Rev.\ Mod.\ Phys.\ }{{\bf #1} {(#2)} {#3}}}
\nc{\rpp}[3]{{\it Rep.\ Prog.\ Phys.\ }{{\bf #1} {(#2)} {#3}}}
\nc{\sjnp}[3]{{\it Sov.\ J.\ Nucl.\ Phys.\ }{{\bf #1} {(#2)} {#3}}}
\nc{\spjetp}[3]{{\it Sov.\ Phys.\ JETP\ }{{\bf #1} {(#2)} {#3}}}
\nc{\yf}[3]{{\it Yad.\ Fiz.\ }{{\bf #1} {(#2)} {#3}}}
\nc{\zetp}[3]{{\it Zh.\ Eksp.\ Teor.\ Fiz.\  }{{\bf #1}  {(#2)} {#3}}}
\nc{\zp}[3]{{\it Z.\ Phys.\ }{{\bf #1} {(#2)} {#3}}}
\nc{\ibid}[3]{{\sl ibid.\ }{{\bf #1} {#2} {#3}}}
\nc{\rf}[1]{(\ref{#1})}
\nc{\nn}{\nonumber \\*}
\nc{\bfB}{\bf{B}}
\nc{\bfv}{\bf{v}}
\nc{\bfx}{\bf{x}}
\nc{\bfy}{\bf{y}}
\nc{\vx}{\vec{x}}
\nc{\vy}{\vec{y}}
\nc{\oB}{\overline{B}}
\nc{\oI}{\overline{I}}
\nc{\oR}{\overline{R}}
\nc{\rar}{\rightarrow}
\nc{\ti}{\times}
\nc{\slsh}{\hskip-5pt/}
\nc{\sm}{Standard~Model~}
\nc{\MP}{M_{\rm Pl}}
\nc{\tp}{t_{\rm Pl}}
\nc{\ave}{\bar{E}}

\nc{\eff}{{\rm eff}}
\nc{\kk}{\vek{k}}
\nc{\pp}{{\rm p}}
\nc{\ga}{g_{a\gamma}}
\nc{\vv}{\\}
\nc{\eee}{{\bf E}}
\nc{\bbb}{{\bf B}}
\nc{\qcd}{T_{\rm QCD}}
\nc{\G}{\rm \ G}
\def\vec#1{{\bf #1}}

\def\lae{\;^{<}_{\sim} \;} \def\gae{\; ^{>}_{\sim} \;} 

\def\ell{e^{c}LL}

\begin{document}
{\title{\vskip-2truecm{\hfill {{\small \\
	\hfill \\
	}}\vskip 1truecm}
{\LARGE Cosmological Constraints on Unparticles as Continuous Mass Particles
}}
{\author{
{\sc \large John McDonald $^{1}$}\\
{\sl\small Cosmology and Astroparticle Physics Group, University of Lancaster,
Lancaster LA1 4YB, UK} }
\maketitle
\begin{abstract}
\noindent

   We study the cosmological constraints on unparticle interactions and the temperature of the Universe for the case where unparticle states can be modelled as continuous mass particles with lifetime $\gae 1$s. By considering thermal background quark decay to continuous mass scalars via a scalar operator of dimension $d_{U}$, we show that the condition that the Universe is not dominated by scalars at nucleosynthesis imposes a lower bound on the scale of the interaction of the unparticle sector, with $M_{U} \gae 20-2600 \TeV$ for $\Lambda_{U} \gae 1 \TeV$, $1.1 \leq d_{U} \leq 2.0$ and $2 \leq d_{BZ} \leq 4$. The existence of a long-lived scalar sector also imposes an upper bound on the temperature of the Universe during radiation-domination, which can be as low as a TeV for $M_{U}$ close to its lower bound.


\end{abstract} 
\vfil
 \footnoterule{\small  $^1$j.mcdonald@lancaster.ac.uk}   
 \newpage 
\setcounter{page}{1}                   

\section{Introduction}

    Recently there has been much interest in the possible existence of a conformally-invariant hidden sector consisting of 'unparticles' \cite{up1,up2} \footnote{Similar ideas were proposed earlier in \cite{vdb}.}; see [4-11]. 
In this picture, the conformally-invariant unparticle sector is generated non-perturbatively from a Banks-Zaks (BZ) sector consisting of non-Abelian gauge fields and vector pairs of massless fermions \cite{bz}, which are assumed to interact very weakly with the SM sector via operators suppressed by a mass scale $M_{U}$. 
The BZ sector undergoes dimensional transmutation to the unparticle phase, corresponding to a strongly-coupled conformal field theory (CFT), below an infra-red fixed point energy $\Lambda_{U}$. This unparticle phase has no conventional particle description but interacts with the SM via unparticle operators of mass dimension $d_{U}$, which create unparticle states. The remarkable characteristic of unparticles is that the phase space in decay processes to unparticle stuff is the same as the phase space for decay to $d_{U}$ massless particles, where $d_{U}$ can be non-integral. This could provide a distinctive signal for unparticle production in collider experiments in the case where the unparticle description is valid at future collider energies, corresponding to $\Lambda_{U} \gae 1 \TeV$. 

    The couplings of the unparticle operators and the value of $d_{U}$ will depend on the underlying BZ operators and the infra-red dynamics of dimensional transmutation. Since these are model-dependent, most studies of unparticle physics have introduced several possible unparticle operators with unknown coefficients and investigated their consequences. Under reasonable assumptions regarding the unknowns, the mass scale $M_{U}$ can then be constrained by the requirement that the unparticle sector does not conflict with present phenomenological \cite{upheno}, astrophysical \cite{uastro}, long-range force \cite{lrf} and cosmological \cite{dav} observations. 

    Here we consider the effect of an unparticle sector on the cosmology of the SM and the resulting constraints on unparticle interactions and the temperature of the Universe. This requires a model 
for the unparticle states and their interactions with the SM.  
In \cite{steph}, the production of unparticle states by scattering of SM particles was modelled by scalar particles with a continuous mass. (See also \cite{krasnikov,niko}.) Although this picture does not fully describe the strongly-coupled unparticle CFT, we will consider it as a starting point from which to study the cosmology of the unparticle states. 
Therefore in the following we will study the cosmology of continuous mass scalars as motivated by unparticles.  Our focus will be the decay and annihilation rates of thermal background SM quarks to unparticle-based continuous mass scalars. For clarity, we focus on a specific interaction between SM quarks and a scalar unparticle operator, leaving a more general operator analysis for future work.

   Our analysis is based on the assumption that the cosmology of continuous mass scalars will approximate that of unparticles. It should be emphasized that this is a non-trivial assumption, which can only be validated with a deeper study of unparticle cosmology based on a full CFT analysis. Continuous mass scalars imply the existence of asymptotic states, which do not formally exist in a CFT.  Therefore continuous mass scalars cannot fully describe the physics of unparticles. The degree to which they correctly describe unparticle cosmology is therefore dependent on the degree to which continuous mass scalars accurately describe the production of unparticle states in particle physics processes and the stability of the unparticle energy density. 

    The scale-invariance of the unparticles must be broken in order that they are not overproduced in astrophysical processes, which requires that there is a mass gap larger than 30 MeV \cite{uastro}. The unparticle 
description is valid for unparticle production occuring at energy scales larger than the mass gap, otherwise a standard particle cosmology applies. We therefore assume that the mass gap is sufficiently small that there is a range of temperature between the mass gap and $\Lambda_{U}$ for which unparticle states are produced.   

           The article is organised as follows. In Section 2 we review unparticles and their interpretation in terms of deconstructing scalars. We also relate these scalars to a physical interpretation of the unparticle states. In Section 3 we calculate the decay and pair annihilation rates of thermal SM quarks to unparticle-based continuous mass scalars and derive bounds from requiring that the scalar energy density does not disturb nucleosynthesis. In Section 4 we discuss our conclusions.

\section{The Model} 

              The unparticle model \cite{up1,up2} is based on an interaction between Banks-Zaks (BZ) fields and SM fields 
mediated by exchange of heavy particles of mass $M_{U}$ 
\be{e1}        \frac{1}{M_{U}^{k}}O_{SM}O_{BZ}               ~.\ee 
$O_{BZ}$ is an operator with mass dimension $d_{BZ}$ made out of BZ fields. At energy scales less than $\Lambda_{U}$ the 
BZ sector undergoes dimensional transmutation and a scale-invariant unparticle sector, corresponding to a strongly self-coupled CFT, is formed. The BZ operators match onto unparticle operators and \eq{e1} becomes
\be{e2}         \frac{C_{U} \Lambda_{U}^{d_{BZ} - d_{U}} }{M_{U}^{k}} O_{SM} O_{U}      ~,\ee 
where $d_{U}$ is the scaling dimension of the unparticle operators $O_{U}$. The value of $C_{U}$ and $d_{U}$ 
will be determined by the infra-red dynamics of the Banks-Zaks fields and the form of $O_{BZ}$. These will be considered as free parameters in the following.  
To study how the unparticle sector influences the cosmology of the SM, we focus on a specific interaction between SM quarks and a scalar unparticle operator $O_{U}$ \cite{up1},  
\be{e3}  \frac{i \lambda}{\Lambda_{U}^{d_{U}}} \overline{q} \gamma_{\mu}\left(1- \gamma_{5}\right) q^{'} \partial^{\mu} 
O_{U} + h.c.     ~.\ee 
Here $\lambda \equiv C_{U} (\Lambda_{U}/M_{U})^{d_{BZ}}$ is a dimensionless coupling. 

   A useful conceptual framework for understanding unparticle production was presented in \cite{steph}. It was shown that by breaking scale-invariance in a controlled way, the continuous energy spectrum of the unparticle sector can be replaced by a discrete tower of deconstructing scalar particles $\phi_{n}$ of mass $M_{n}$, with mass spacing controlled by a mass parameter $\Delta$ such that $M_{n}^{2} = n \Delta^{2}$. In this case the unparticle operator $O_{U}$ is replaced by a sum over canonically normalized deconstructing scalar fields $\phi_{n}$, 
\be{e4}  \sum_{n = 1}^{\infty} \frac{i \lambda F_{n}}{\Lambda_{U}^{d_{U}}} \overline{q} \gamma_{\mu}\left(1- \gamma_{5}\right) q^{'} \partial^{\mu} \phi_{n} + h.c.     ~,\ee 
where 
$$    F_{n}^{2} =  \frac{A_{d_{U}}}{2 \pi} \Delta^{2} (M_{n}^{2})^{d_{U} - 2}      ~$$
and 
$$A_{d_{U}} =  \frac{16 \pi^{5/2}}{\left(2 \pi\right)^{2 d_{U}}} \frac{\Gamma\left(d_{U} + 1/2\right)}{\Gamma\left(d_{U} - 1\right)\Gamma\left(2 d_{U}\right) }      ~.$$
$A_{d_{U}}$ is a conventionally chosen phase space factor for unparticles \cite{up1}; only the combination $C_{U} 
A_{d_{U}}^{2}$ appears in physical processes. In the limit $\Delta \rightarrow 0$, the sum over decays $q^{'} \rightarrow  q + \phi_{n}$ for all kinematically allowed values of $n$ produces an expression for $d\Gamma/d E_{q}$ in agreement with the direct unparticle calculation of $q^{'} \rightarrow q + U$ based on $O_{U}$ \cite{steph}. The individual $q + \phi_{n}$ final states of energy $E_{q} = (m_{q^{'}}^{2} - M_{n}^{2})/2m_{q^{'}}$ merge into a continuum decay rate as a function of $E_{q}$ as $\Delta \rightarrow 0$, reproducing the unparticle result.

        The scalar particle deconstruction of the unparticle propagator 
does not completely describe the unparticle CFT. However, in cosmology we are primarily interested in the scattering between SM particles and unparticle states. In this case the continuous mass scalars may 
be a useful approximation to the 
unparticle states. If we consider the unparticle sector to be the strongly-coupled limit of a Banks-Zaks 
sector, then the states will correspond to 
`bound states' of Banks-Zaks fermions and gauge fields. Since the theory is scale-invariant in the strongly-coupled limit, there can be no preferred mass associated with the bound states. These objects will therefore have all kinematically possible masses as a result of the scale-invariance of the underlying CFT. 
Therefore the unparticle states may be expected to behave approximately as 
particle-like states with a continuous mass parameter. 
This picture cannot be exactly correct however, as CFTs do not have asymptotic states. In what follows we are therefore assuming that continuous mass particles are a sufficiently good approximation to the unparticle states that they can be used to understand the production and stability of an unparticle energy in a cosmological setting.     
Deviations from the unparticle picture may also be expected due to the finite value of $\Lambda_{U}$ relative to the interaction energy in scattering processes, which breaks true scale-invariance.

\section{Cosmological constraints from quark decay and annihilation to continuous mass scalars} 

    We will consider the couplings \eq{e3} and \eq{e4} in the context of cosmology. In particular, we will calculate the decay and annihilation rates of thermal background quarks to continuous mass scalars to obtain the conditions under which these processes can produce a significant energy density and the resulting constraints on cosmology. 

         A basic property of the continuous mass particle sector from unparticle deconstruction is that the particles are stable \cite{steph}.  
This can be understood by considering the deconstructing scalars in the limit $\Delta \rightarrow 0$. In this limit the couplings of the scalars,
which are proportional to $F_{n} \propto \Delta ^{d_{U} -1}$, tend to zero if $d_{U} > 1$. The reason the decay and scattering rates of SM particles to continuous mass scalars is finite in the $\Delta \rightarrow 0$ limit is that the decrease in the rate to a scalar particle final state $\phi_{n}$ is compensated for by an increase in the number of kinematically allowed scalar final states $\approx E^{2}/\Delta^{2}$, where $E$ is the energy the decaying or annihilating quark. (We will demonstrate this explicitly for the case of pair annihilation of SM quarks to continuous mass scalars.) However, if we consider the decay and scattering of continuous mass scalars to SM particles, then as $\Delta \rightarrow 0$ there is no compensating increase in the number of SM particles. So the rates of all processes which could result in the decay of the scalar particle energy density to SM particles tend to zero in the continuous mass limit $\Delta \rightarrow 0$ \cite{steph}. Since the cosmological constraints on unparticles follow from their effect on nucleosynthesis, unstable unparticles will be subject to the same constraints if their lifetime is greater than O(1)s.

        Since an unparticle sector is scale-invariant, the unparticle density evolves with scale factor as $\rho_{U} \propto a^{-4}$ \cite{dav}. We will assume the continuous mass scalar density also evolves as $a^{-4}$. (If it evolves as $a^{-n}$ with $n < 4$ then the cosmological constraints become stronger.) If the scalar density decouples from the SM density at $T_{DEC} \gg T_{BBN} \approx 0.1 \MeV$, where $T_{BBN}$ is the temperature at Big-Bang Nucleosynthesis (BBN), then a dilution of the continuous mass scalar density relative to the radiation density is possible, due to subsequent photon heating by annihilation of massive SM particles. However, the maximum dilution relative to photons is only by a factor $\left(g_{*}(T_{BBN})/g_{*}(T_{DEC})\right)^{4/3} \left(g(T_{DEC})/g(T_{BBN})\right)$, where $g$ and $g_{*}$ are the effective number of relativistic degrees of freedom associated with energy and entropy respectively, since the radiation energy density is proportional to $g(T)T^4$ while the continuous mass scalar density redshifts as $a^{-4}$ with $g_{*}(T)a^3 T^3 = constant$. The maximum dilution is obtained if decoupling occurs at a higher temperature than the electroweak phase transition, in which case $g(T_{DEC}) = g_{*}(T_{DEC})= 106.75$. With $g(T_{BBN}) = 3.36$ and $g_{*}(T_{BBN}) = 3.90$ for photons plus neutrinos, this gives dilution by at most a factor 0.39. This is not enough to dilute the continuous mass scalar density sufficiently unless $\rho_{U} < 0.15 \rho_{SM}$ at $T_{DEC}$, since successful nucleosynthesis requires that any additional energy density be less than 6$\%$ of the total at $T_{BBN}$ \cite{bbnlimit}. Therefore in order that the successful light element abundance predictions of BBN are unaffected, SM quark decay or pair annihilation to continuous mass scalars should not produce a large continuous mass scalar density. 

\subsection{Quark decay to continuous mass scalars} 

      The interaction \eq{e4} implies a differential decay rate to deconstructing scalars (equal to the decay rate to unparticles) $q^{'} \rightarrow q + U$ given by \cite{up1,steph}
\be{e5}  \frac{d \Gamma}{d E_{q}} = \frac{ |\lambda|^{2}}{2 \pi^{2}} \frac{m_{q^{'}}^{2} A_{d_{U}} E_{q}^{2} \left(m_{q^{'}}^{2} - 2 m_{q^{'}} E_{q} \right)^{d_{u}-2}  }{\Lambda_{u}^{2 d_{U}}}   ~.\ee
This is the decay rate in the $q^{'}$ quark rest frame. Integrating this for all energies up to $m_{q^{'}}/2$ gives 
\be{e6} \Gamma = \frac{ |\lambda|^{2}}{8 \pi^{2}} \frac{ A_{d_{U}} m_{q^{'}}^{2 d_{U} + 1} }{ d_{U}(d_{U}^{2} - 1)\Lambda_{U}^{2 d_{U}}}   
~.\ee 
In the limit $d_{U} \rightarrow 1$ the decay rate becomes infinite, therefore $d_{U} > 1$ must be imposed \cite{up1}, in agreement with general unitarity arguments for a scalar unparticle operator. 
For thermal relativistic quarks of mean energy $E \approx 3T$, the decay rate $\Gamma_{d}$ is obtained from the rest frame decay rate by dividing by the Lorentz factor $\gamma = 3T/m_{q^{'}}$. Therefore 
\be{e7} \Gamma_{d} \approx \frac{ |\lambda|^{2}}{8 \pi^{2}} \frac{ A_{d_{U}} }{d_{U}(d_{U}^{2} - 1) \Lambda_{U}^{2 d_{U}}}  
\frac{m_{q^{'}}^{2 d_{U} + 2} }{3T}    
~.\ee 
At $T < T_{EW}$, where $T_{EW}$ is the temperature of the electroweak phase transition, we will assume that the Higgs expectation value is given by the $T = 0$ value\footnote{The Higgs expectation value is temperature dependent, given by 
$<H> = v(1-T^2/T_{EW}^2)^{1/2}$, where $T_{EW} \approx 1.2 m_{h}$ \cite{hall}. 
For simplicity in our estimates we will use the approximation that \newline $<H> = 0$ at $T < T_{EW}$ and $<H> = v$ at $T > T_{EW}$.} $<H> = v$. 
At $T > T_{EW}$, we expect the effective mass of the quarks to be given by the temperature, 
$m_{q^{'}} \approx T$. 
Therefore 
\be{e9a} \Gamma_{d} \approx \frac{ C_{U}^{2} A_{d_{U}} }{24 \pi^{2}}
 \frac{ \Lambda_{U}^{2(d_{BZ} - d_{U})} }{d_{U}(d_{U}^{2} - 1)M_{u}^{2 d_{BZ}}}  
\frac{m_{q^{'}}^{2 (d_{U} + 1)} }{T}    \;\;\;\;  ;   \;\;  T < T_{EW}
~\ee 
and
\be{e9b} \Gamma_{d} \approx \frac{ C_{U}^{2} A_{d_{U}} }{24 \pi^{2}}
 \frac{ \Lambda_{U}^{2(d_{BZ} - d_{U})} T^{2 d_{U} + 1}    }{d_{U}(d_{U}^{2} - 1)M_{u}^{2 d_{BZ}}}  
\;\;\;\;  ;   \;\;  T > T_{EW}
~.\ee 
In order not to produce a large continuous mass scalar density we must then impose that 
$\Gamma_{d} < H$, where during SM radiation-domiation $H = k_{T}T^2/M$ 
with $k_{T} = (\pi^{2}g(T)/90)^{1/2}$ and $M = M_{Pl}/\sqrt{8 \pi}$. 

   For $T < T_{EW}$, the strongest constraint comes from the case of t quark decay. In this case 
the condition $\Gamma_{d} < H$ is strongest at the lowest value of $T$ for which there are relativistic thermal t-quarks, 
$T \approx m_{t}/3$, which implies that
\be{e10}   M_{U}^{2 d_{BZ}}  \gae \frac{9}{8 \pi^2} \frac{k_{1}}{k_{T}}   
\frac{m_{t}^{2 d_{U}-1}M\Lambda_{U}^{2(d_{BZ} - d_{U})}}{d_{U}(d_{U}^{2} - 1)}  
~,\ee
where $k_{1} =  C_{U}^{2} A_{d_{U}} $. If this condition is satisfied, then for $m_{t}/3 \lae T < T_{EW}$ 
there is no significant decay of the SM radiation to continuous mass scalars.  Once $T \lae m_{t}/3$, the next possible decay is to b quarks, but the condition for this not to occur, which replaces $m_{t}$ by $m_{b}$ in \eq{e10}, is automatically satisfied if \eq{e10} is satisfied\footnote{Decay to other light quarks is also possible, but these decays will be subdominant due to the quark mass factor in \eq{e10}.}. On the other hand, if \eq{e10} is not satisfied, then as $T$ decreases from $T_{EW}$, at some $T \gae m_{t}/3$ the decay rate becomes faster than $H$. The SM radiation then loses energy to continuous mass scalars, decreasing $T$ and increasing $\Gamma_{d}$ relative to $H$ (which remains constant throughout since the total energy density is constant) until $T \lae m_{t}/3$. As a result, if $T$ is larger than $m_{t}/3$, a large fraction of the SM radiation energy will be transferred to the continuous mass scalar sector, resulting in a scalar-dominated Universe.   

    For $T > T_{EW}$, the condition $\Gamma_{d} < H$ requires that 
\be{e11}   T^{2 d_{U} - 1} \lae 24 \pi^{2}
\frac{k_{T}}{k_{1}} 
 \frac{ d_{U}(d_{U}^{2} - 1) M_{u}^{2 d_{BZ}}}{ \Lambda_{U}^{2(d_{BZ} - d_{U})}M }    ~.\ee 
If this condition is not satisfied, then SM radiation will rapidly decay to continuous mass scalars until
$T$ has decreased sufficiently that the condition $\Gamma_{d} < H$ is satisfied. As a result, a large fraction of the SM radiation energy will have transfered to the scalar sector when the decay rate becomes negligible. 
Therefore there is an upper bound on the temperature of the radiation-dominated universe in the presence of a sufficiently stable (lifetime $\gae 1$s) continuous mass scalar sector.

     These conclusions are based on the idea that any process which 
results in energy transfer will always be much more rapid from the SM to the continuous mass scalar sector than the reverse process. In the deconstructing scalar picture the unparticle states are stable with 
respect to decay and annihilation via the interaction \eq{e3}. 
However, it is conceivable that unparticle self-interactions, which 
will depend on the details of the underlying CFT, could results in processes transferring energy to the SM. Nevertheless, the reverse process from SM to continuous mass scalars will always be much more rapid due to the large number of scalar particle final states. 

           We note that if the unparticle CFT can come into thermal  equilibrium then it may be possible to evade these constraints, if the central charge $c$ of the CFT is small enough. The energy density is 
$\rho_{U} = \pi c T^{4}/6$. Therefore if the unparticle density 
decouples at $T \gae M_{W}$, the density at nucleosynthesis will be small enough if $ \pi c/6 < 0.15 \times \pi^{2} g(T)/30$ with $g(T) =106.75$. This requires $c < 10$. A considerable reduction in the number of effective degrees of freedom,  
from $100$ for the underlying Banks-Zaks theory to less than 10 for the unparticle CFT, would be necessary for this to work.

\subsection{Quark pair annihilation to continuous mass scalars}

      We next consider the annihilation of thermal quark pairs to continuous mass scalars, by calculating the 
scattering cross-section to deconstructing scalars in the $\Delta \rightarrow 0$ limit. The CM 
annihilation cross-section $q + q \rightarrow \phi_{m} + \phi_{n}$ following from \eq{e4} is given by 
\be{e12}   \sigma_{mn} \approx   \frac{g_{m}^2 g_{n}^2 E^2}{M_{s}^{4}}    ~,\ee
where $E$ is the energy of the annihilating quarks, 
$$ g_{n}^{2} = \frac{|\lambda|^{2}}{n^{(2-d_{U})}}    $$
and 
$$  M_{s}^{2} = \frac{2 \pi}{A_{d_{U}}} \left( \frac{\Lambda_{U}}{\Delta} \right)^{2 d_{U}} \Delta^{2}    ~.$$ 
(In this we have written the coupling for each $n$ in \eq{e4} as $g_{n}/M_{s}$.) 
The total scattering rate to deconstructing scalars is obtained by summing over the kinematically allowed final
states. With quark energy $E$, the final state scalars have $n_{max} = E^2/\Delta^2$, therefore 
\be{e13}    \sigma_{TOT} =  \sum_{n = 1}^{n_{max}} \sum_{m = 1}^{m_{max}} \sigma_{mn}     ~.\ee
For $n_{max} \gg 1$ we can replace the sums with integrals, 
\be{e14}  \sum_{n = 1}^{n_{max}} g_{n}^{2} \approx \int_{1}^{E^{2}/\Delta^{2}} g_{n}^{2} dn 
\equiv \int_{1}^{E^{2}/\Delta^{2}}   \frac{|\lambda|^{2}}{n^{(2-d_{U})}} dn     ~.\ee 
Therefore
\be{e15}  \sum_{n = 1}^{n_{max}} g_{n}^{2}  
\approx \frac{|\lambda|^2}{\left(d_{U} - 1\right)} \left[ \left(\frac{E}{\Delta}\right)^{2\left(d_{U}-1\right)} - 1 \right]     ~.\ee 
Thus summing over $m$ and $n$ is \eq{e13} gives 
\be{e16}   \sigma_{TOT} \approx \frac{E^{2}}{ M_{s}^{4}}  \frac{|\lambda|^4}{\left(d_{U} - 1\right)^{2}} \left[ \left(\frac{E}{\Delta}\right)^{2\left(d_{U}-1\right)} - 1 \right]^{2}     ~.\ee
Replacing $M_{s}^{2}$ by its definition then gives 
\be{e17} \sigma_{TOT} \approx \frac{|\lambda|^4 E^2}{\left(d_{U} - 1\right)^{2}} \left[ \left(\frac{E}{\Delta}\right)^{2\left(d_{U}-1\right)} - 1 \right]^{2}   \frac{A_{d_{U}}^{2} \Delta^{4\left(d_{U} - 1\right)} }{
(2 \pi)^2 \Lambda_{U}^{4 d_{U}}    }       ~.\ee     
In the case $d_{U} > 1$, in the continuous mass limit $\Delta \rightarrow 0$ the cross-section tends to a finite limit 
\be{e18} \sigma_{TOT} \approx \frac{|\lambda|^{4}}{\left(d_{U} - 1\right)^{2}} 
 \frac{A_{d_{U}}^{2} E^{4 d_{U} - 2}}{(2 \pi)^{2} \Lambda_{U}^{4 d_{U}}  }    ~.\ee 
In the case $d_{U} \leq 1$, $\sigma_{TOT} \rightarrow \infty$ as $\Delta \rightarrow 0$. Therefore, as in the case of the quark decay rate, the quark annihilation rate is singular as $d_{U} \rightarrow 1$. 
 
       For $d_{U} > 1$ the annihilation rate is then $\Gamma_{ann} = n_{q} \sigma_{TOT}$, where 
$n_{q} = 12T^{3}/\pi^{2}$ is the number density of thermal quarks $q$. Therefore, with $E \approx 3T$, 
\be{e19}   \Gamma_{ann} \approx  \frac{1}{9 \pi^{4}} \frac{C_{U}^{4} A_{d_{U}}^2}{ \left(d_{U} - 1\right)^{2} }
\frac{ (3T)^{4 d_{U} +1}}{\Lambda_{U}^{4 d_{U}} }    ~.\ee
In most cases this rate is small compared with the quark decay rate. 
However, for $d_{U}$ sufficiently close to 1, the annihilation rate 
($\propto (d_{U} - 1)^{-2}$) can become larger than the decay rate \eq{e7} ($\propto (d_{U} - 1)^{-1}$).    

\subsection{Cosmological constraints on unparticle parameters from continuous mass scalar cosmology} 

            We next consider the constraints on the unparticle parameters $M_{U}$ and 
$\Lambda_{U}$ as a function of $d_{U}$ following from 
thermal quark decay to continuous mass scalars. The bounds derived above assume that the energy of the quarks is less than 
$\Lambda_{U}$, otherwise the decay and annihilation rates should be calculated for BZ particle final states. Since the most interesting case is where unparticles may be observed in future colliders, we will consider $\Lambda_{U} \gae 1 \TeV$ in the following. The decay rate depends upon the product $k_{1} \equiv C_{U}^{2} A_{d_{U}}$, which is determined by the infra-red dynamics of the Banks-Zaks fields \cite{up1}. Therefore any conclusions about the cosmology of unparticles is dependent upon assumptions about this product, which should become better understood as unparticle physics develops. In the following we will assume that $k_{1}$ is not very large or small compared with 1. 

      The constraints will depend on the assumed values of the model parameters $d_{BZ}$, $d_{U}$ and $\Lambda_{U}$. We first focus on the  case where the BZ operator is a quark bilinear, so that $d_{BZ} = 3$, and calculate constraints for the cases $d_{U} = 1.1, \; 1.5$ and 2 when $\Lambda_{U} \gae 1 \TeV$. 
              
  The most important constraints come from \eq{e10} when $T < T_{EW}$. This gives lower bounds on $M_{U}$ as a function of $d_{U}$ 
$$  {\bf d_{U} = 1.1:} \;\;\;\;  M_{U} \gae 190\; k_{1}^{1/6} 
\left(\frac{\Lambda_{U}}{1 \TeV}\right)^{19/30} \TeV      $$ 
$$  {\bf d_{U} = 1.5:} \;\;\;\;  M_{U} \gae 110 \; k_{1}^{1/6} 
\left(\frac{\Lambda_{U}}{1 \TeV}\right)^{1/2} \TeV      $$ 
\be{e22}  {\bf d_{U} = 2.0:} \;\;\;\;  M_{U} \gae 70 \; k_{1}^{1/6} 
\left(\frac{\Lambda_{U}}{1 \TeV}\right)^{1/3}  \TeV      ~,\ee
where we have used 
$k_{T} = 3$. 
Thus if $k_{1}$ is not very large or small compared with 1 and $\Lambda_{U} \gae 1 \TeV$, 
the mass scale of the $d_{BZ} = 3$ interaction between the BZ and SM fields must be greater than around 100 TeV for $d_{U}$ between 1.1 and 2.0. These bounds hold if the radiation-dominated era has $T \gae m_{t}/3$ at some time, otherwise weaker bounds based on lighter quark decays will apply. 

   Upper bounds on the temperature of the radiation-dominated era when $T > T_{EW}$ can be obtained from \eq{e11}. However, these bounds stricly hold only if the energy of the decaying quark $E \approx 3T$ is less than  $\Lambda_{U}$, since for larger temperatures and energies the decay will produce BZ particles rather than unparticles. 
For $T_{EW} < T \lae \Lambda_{U}/3$ we obtain:
$$  {\bf d_{U} = 1.1:} \;\;\;\;  T \lae 10 \TeV 
\left(\frac{1}{k_{1}}\right)^{5/6} 
\left(\frac{M_{U}}{250 \TeV}\right)^{5}     
\left(\frac{1 \TeV}{\Lambda_{U}}\right)^{19/6}    $$ 
$$  {\bf d_{U} = 1.5:} \;\;\;\;  T \lae 2.6 \TeV
\left(\frac{1}{k_{1}}\right)^{1/2} 
\left(\frac{M_{U}}{150 \TeV}\right)^{3}     
\left(\frac{1 \TeV}{\Lambda_{U}}\right)^{3/2}    $$ 
\be{e25}  {\bf d_{U} = 2.0:} \;\;\;\;  T \lae 1.3 \TeV 
\left(\frac{1}{k_{1}}\right)^{1/3} 
\left(\frac{M_{U}}{100 \TeV}\right)^{2}     
\left(\frac{1 \TeV}{\Lambda_{U}}\right)^{2/3}    ~.\ee 
The upper bounds on $T$ are sensitive to $M_{U}$ and the value of $3T$ can easily be above $\Lambda_{U}$. In this case the decay should be calculated to BZ sector fields using the interaction \eq{e1}. For $d_{BZ} = 3$ and $O_{BZ}$ corresponding to a bilinear of BZ fermions $\overline{\Psi}\Psi$, the interaction with the SM quarks becomes 
\be{e26}        \frac{1}{M_{U}^{k}}O_{SM}O_{BZ}  \rightarrow \frac{ \partial_{\mu}(\overline{q} \gamma_{\mu}\left(1- \gamma_{5}\right) q^{'} ) \overline{\Psi}\Psi }{M_{U}^{3}}              ~.\ee 
The decay rate for $q^{'} \rightarrow q \overline{\Psi} \Psi$ is then 
\be{e27} \Gamma_{d} \approx \frac{1}{\left(8 \pi\right)^3} \frac{T^{7}}{M_{U}^{6}}            ~\ee
The condition $\Gamma_{d} < H$ then implies that 
\be{e28} T \; \lae \; 1.8 \; \left(\frac{M_{U}}{100 \TeV}\right)^{6/5} \TeV    ~.\ee 
If this bound is not satisfied, energy will flow to the BZ sector until 
thermal equilibrium is established with the BZ sector, 
resulting in a large BZ sector energy density. 
For example, for the case of the original Banks-Zaks theory with an $SU(3)$ gauge group and $16.5 > N_{F} > 8$ Dirac fermions transforming in the 
fundamental representation of $SU(3)$ \cite{bz}, the number of thermal degrees of freedom is at least $g(T) = 8 \times 2 + 7/8 \times 4 \times 3 \times 8 = 100.0$, which is similar to the number of degrees of freedom in the SM, $g(T) = 106.75$. Therefore in thermal equilibrium, $\rho_{BZ} \approx \rho_{SM}$. Once $T$ drops below the upper limit in \eq{e28}, the SM and BZ sectors decouple. As the BZ fields lose energy via expansion they will evolve into continuous mass scalars, leaving an energy density in the stable scalar sector which is too large to dilute sufficiently by photon heating in the SM before nucleosynthesis.
Therefore, a stable unparticle-based continuous mass scalar sector in general implies an upper limit on the temperature of the radiation-dominated Universe, which is of the order of 1 TeV for $M_{U}$ close to its lower limit.

       The above results are for the case $d_{BZ} = 3$. To show the sensitivity to $d_{BZ}$ we calcuate the lower bound on 
$M_{U}$ from \eq{e10} for the cases $d_{BZ} = 2$ and 4:
\newline \underline{$d_{BZ} = 2$} 
$$  {\bf d_{U} = 1.1:} \;\;\;\;  M_{U} \gae 2600 \; k_{1}^{1/4} \left(\frac{\Lambda_{U}}{1 \TeV}\right)^{9/20} \TeV   $$ 
$$  {\bf d_{U} = 1.5:} \;\;\;\;  M_{U} \gae  1100 \; k_{1}^{1/4} 
\left(\frac{\Lambda_{U}}{1 \TeV}\right)^{1/4} \TeV      $$ 
\be{e29a}  {\bf d_{U} = 2.0:} \;\;\;\;  M_{U} \gae 520 \; k_{1}^{1/4}  \TeV\; . \;\;\;\;\;\;\;\;\;\;\;\;\;\;\;\;\;\;     
 ~\ee 
\underline{$d_{BZ} = 4$}
$$  {\bf d_{U} = 1.1:} \;\;\;\;  M_{U} \gae 43 \; k_{1}^{1/8} \left(\frac{\Lambda_{U}}{1 \TeV}\right)^{29/40}  \TeV   $$ 
$$  {\bf d_{U} = 1.5:} \;\;\;\;  M_{U} \gae  33 \; k_{1}^{1/8} 
\left(\frac{\Lambda_{U}}{1 \TeV}\right)^{5/8} \TeV      $$ 
\be{e30a}  {\bf d_{U} = 2.0:} \;\;\;\;  M_{U} \gae 23 \; k_{1}^{1/8} 
\left(\frac{\Lambda_{U}}{1 \TeV}\right)^{1/2} \TeV       .\ee 
Larger $d_{BZ}$ for a given $d_{U}$ reduces the lower bound on $M_{U}$ for $\Lambda_{U} \approx 1 \TeV$, but increases the sensitivity to $\Lambda_{U}$. From \eq{e22}, \eq{e29a} and \eq{e30a} we see that  
the lower bounds on $M_{U}$ are in the range $20 \TeV$ to $2600 \TeV$ when $1.1 \leq d_{U} \leq 2.0$ and $2 \leq d_{BZ} \leq 4$, assuming that $\Lambda_{U} \gae 1 \TeV$ and $k_{1}$ is not too different from 1.  

\subsection{Discussion}

     The first study of unparticle cosmology was presented in 
\cite{dav}. This was based on a purely dimensional estimate of the rate of production $\Gamma_{\psi}$ of unparticles from SM radiation via a vector unparticle operator interaction of the form $\overline{\psi} \gamma_{\mu} \psi O_{U}^{\mu}$. It was assumed that once $\Gamma_{\psi} > H$, the unparticles and SM particles are in thermal equilibrium at $T$. Once the unparticles decouple, the unparticle 
sector is assumed to be stable and to have a temperature $T_{U}$ which is less than the photon temperature $T$ due to photon heating by subsequent annihilations. In order to ensure that $T_{U}$ is sufficiently suppressed that the additional unparticle energy does not affect BBN, it was required that decoupling of the unparticles 
occurs at $T \gae 1 \GeV$, before the QCD phase transition. 
Requiring this imposes an upper bound on $\Lambda_{U}$. However, as we have shown, the actual suppression of the unparticle density relative to the radiation density is by at most a factor of 0.39,  which is ineffective in suppressing a decoupled stable unparticle density with $\rho_{U} > 0.15 \rho_{SM}$.

            The main difference with our discussion of unparticle cosmology based on stable continuous mass scalars is in the interpretation of the condition $\Gamma_{d} > H$. Rather than establishing thermal equilibrium between the SM and the continuous mass scalar sector, we interpret this as a rapid flux of energy from SM radiation to the scalar sector, resulting in a scalar-dominated Universe. Therefore $\Gamma_{d} < H$ must hold for all $T$, imposing an upper bound on $T$ during radiation-domination. Our analysis also differs in the interaction we considered, based on a scalar unparticle operator. The decay and annihilation rates in this case differ from a simple dimensional estimate due to the quark mass factors which must also be included. 

         Strong constraints on unparticles can be imposed by supernovae \cite{uastro} and by unparticle-mediated long-range forces \cite{lrf}. However, these constraints can be evaded by slightly by breaking the conformal invariance, for example by giving the deconstructing scalars a mass $> 30 \MeV$ to suppress their creation in supernovae \cite{del}. This can be achieved by coupling the unparticle operators to Higgs scalars, in which case the continuous mass scalars have a lower bound $m_{gap}$ on their masses but continuous masses at $m > m_{gap}$, which preserves their stability. 

    Since we have shown that $M_{U} \gae 100 \TeV$ is typically necessary for compatibility with SM cosmology, it might appear that detection of unparticles in collider experiments is unlikely if continuous mass scalars serve as a good model for unparticle states. However, in the case where unparticle operators couple to the Higgs doublet, it is possible to detect unparticles with such large $M_{U}$ e.g. by observing the partial Higgs boson decay width to gluon and photon pairs \cite{ko}. 

\section{Conclusions} 

            We have considered unparticle cosmology 
under the assumption that unparticle states can be modelled by continuous mass scalar particles. We have shown that the requirement that sufficiently stable (lifetime $\gae 1$s) continuous mass scalars do not dominate the energy density at nucleosynthesis implies that there is a lower bound on the scale $M_{U}$ of the interaction between the Banks-Zaks fields and the Standard Model and an upper bound on the temperature of the radiation-dominated Universe. 

   For an interaction between SM quarks and scalar unparticle operators with $d_{BZ} = 3$ and $\Lambda_{U} \gae 1 \TeV$, the lower bound on $M_{U}$ is typically of the order of 100 TeV for unparticle dimensions $d_{U}$ in the range 1.1 - 2.0. Varying $d_{BZ}$ widens the range of lower bounds, with $M_{U} \gae 20-2600 \TeV$ for $2 \leq d_{BZ} \leq 4$. Decays will dominate over annihilations except if $d_{U}$ is very close to 1. Including other unparticle operators can only strengthen these bounds. 

      A stable continuous mass scalar sector coupled to the Standard Model implies the existence of an upper bound on the temperature of the Standard Model radiation-dominated era. In the example with $d_{BZ} = 3$, the upper bound is of the order of 1 TeV for $M_{U}$ close to its lower bound. 
 
         These constraints may be evaded if it is possible for the unparticle sector to come into thermal equilibrium initially and to subsequently decouple at $T \gae M_{W}$. Then if the central charge of the unparticle CFT is sufficiently small, $c < 10$, the decoupled unparticle density will be diluted sufficiently to evade nucleosynthesis constraints. This requires a significant reduction in the number of effective degrees of freedom of the unparticle CFT relative to the underlying Banks-Zaks sector, by approximately a factor of 10.

      The significance of these conclusions for unparticle cosmology depends on the degree to which the unparticle states can be approximated by stable continuous mass particles.  We have assumed that for the case of unparticles due to a Banks-Zaks sector with strong-coupling scale $\Lambda_{U}$, the unparticle states may be approximately described by continuous mass particles in the limit where $\Lambda_{U}$ is large compared with the energy of interaction. 
However, this description cannot be exactly correct as CFTs do not have asymptotic states. An full CFT analysis of unparticle cosmology is ultimately necessary to establish the validity of the results suggested by the continuous mass scalar analysis.

In addition, we expect the scale-invariance of the strongly-coupled theory to be broken by the energy of unparticle interactions relative to $\Lambda_{U}$. This could then permit processes allowing the unparticle density to decay or annihilate to SM particles. Cosmological constraints are based on the unparticle density being stable until the time of nucleosynthesis. Therefore if the unparticle density has a lifetime greater than $O(1)$s then the constraints derived in this paper will apply. A recent study \cite{raj} suggested that unparticles have a finite lifetime, whereas in \cite{delstab} it was suggested that unparticle stability is possible, depending on the mass of the pole in the unparticle propagator relative to the mass of the possible SM final states in the decay process. It will be important to establish if unparticles can be sufficiently unstable to evade cosmological constraints from nucleosynthesis.

This work was supported by the European Union through the Marie Curie Research and Training Network "UniverseNet" (MRTN-CT-2006-035863) and by STFC (PPARC) Grant PP/D000394/1.


\end{document}